
%
%
\message{Extended math symbols.}

\ifx\oldzeta\undefined    
  \let\oldzeta=\zeta    
  \def\zzeta{{\raise 2pt\hbox{$\oldzeta$}}} 
  \let\zeta=\zzeta    
\fi

\ifx\oldchi\undefined    
  \let\oldchi=\chi    
  \def\cchi{{\raise 2pt\hbox{$\oldchi$}}} 
  \let\chi=\cchi    
\fi



\def\frac#1#2{{#1 \over #2}}

\def\half{\ifinner {\scriptstyle {1 \over 2}}
   \else {1 \over 2} \fi}

\def\bra#1{\langle#1\vert}  
\def\ket#1{\vert#1\rangle}  

\def\simge{\rlap{\raise 2pt \hbox{$>$}}{\lower 2pt \hbox{$\sim$}}}
\def\simle{\rlap{\raise 2pt \hbox{$<$}}{\lower 2pt \hbox{$\sim$}}}



\def\slashchar#1{\setbox0=\hbox{$#1$}  
   \dimen0=\wd0     
   \setbox1=\hbox{/} \dimen1=\wd1  
   \ifdim\dimen0>\dimen1   
      \rlap{\hbox to \dimen0{\hfil/\hfil}} 
      #1     
   \else     
      \rlap{\hbox to \dimen1{\hfil$#1$\hfil}} 
      /      
   \fi}      %







  \def\za{\alpha}          
  \def\zb{\beta}           
  \def\zd{\delta}          \def\zD{\Delta}
  \def\ze{\epsilon}         
              
  \def\zg{\gamma}

  \def\zl{\lambda}         \def\zL{\Lambda}
  \def\zm{\mu}

  \def\zp{\pi}             
              \def\zQ{\Psi}
              
  \def\zs{\sigma}          \def\zS{\Sigma}
  \def\zt{\tau}
  
            \def\zW{\Omega}


\def\fpi{f_\pi}
\def\mpi{m_\pi}

\def\vpi{\vec\pi}
\def\vtau{\vec\tau}

\def\bra#1{\langle #1 \,\vert}
\def\ket#1{\vert\, #1 \rangle}

\def\dsl{\rlap{/}{\partial}}

\def\wlog#1{} 
\catcode`\@=11

\outer\def\rthnum#1{\topright{RUB-TPII-#1}}

\def\wlog{\immediate\write\m@ne} 
\catcode`\@=12 


\def\AP{\journal{Ann.\ \Phys}}

\def\JPG{\journal{J.\ \Phys}G}

\def\NC{\journal{Nuov.\ Cim.\ }}
\def\NCA{\journalp{Nuov.\ Cim.\ }A}

\def\ZPC{\journalp{Z.\ \Phys}C}
\def\ZPA{\journalp{Z.\ \Phys}A}

\def\APPB{\journalp{Acta \ \Phys \ Polonica}B}

\input phys
\english
\titlepage
\chapters 
\equchap
\equfull
\figpage
\tabpage
\refpage
\overfullrule=0pt
\RF\Egu76{T.Eguchi, {\PRD14(1976)2755*}; D.Ebert and M.K.Volkov,
{\ZPC16(1983)205*}; D.Ebert and H.Reinhardt, \NPA271(1986)188*}
\RF\Ripka84{S.Kahana and G.Ripka, \NPA429(1984)462*}
\RF\Ripka91{G.Ripka and R.M.Galain, Lecture at the XXXCracow School of
Theoretical Physics, June 2-12, 1990, Zkopane, Poland, {\APPB22(1991)187*}
; M.Jaminon, R.M.Galain, G.Ripka and P.Stassart, SPhT/91-037, Preprint
Saclay, March 1991 (submitted to Nucl.Phys.A)}
\RF\Shifman79{M.A.Shifman, A.J.Vainstein and V.I.Zakharov,
{\NPB147(1979)385*}; \NPB163(1980)43*}
\RF\Nam61{Y.Nambu and G.Jona-Lasinio, \PR122(1961)354*}
\RF\Schwinger51{J.Schwinger, \PR82(1951)664*}
\RF\Gel60{M.Gell-Mann and M.Levi, \NC16(1960)705*}
\RF\Fio88{M.Fiolhais, K.Goeke, F.Gr\"ummer and J.N.Urbano,
{\NPA481(1988)727*}; P.Alberto, E.Ruiz Arriola, M.Fiolhais,
F.Gr\"ummer, J.Urbano and K.Goeke, {\PLB208(1988)75*}; P.Alberto,
E.Ruiz Arriola, M.Fiolhais, K.Goeke, F.Gr\"ummer and J.Urbano,
\ZPA336(1990)449*}
\RF\Chr90{Chr.V.Christov, E.Ruiz Arriola and K.Goeke,
{\NPA510(1990)1990*}; {\PLB225(1989)22*}; \PLB243(1990)333*}
\RF\Chr91{Chr.V.Christov, E.Ruiz Arriola and K.Goeke, Lecture presented at
the XXX Cracow School of Theoretical Physics, June 2-12, 1990,
 Zakopane, Poland, \APPB22(1991)187*}
\RF\TMei90{T.Meissner, F.Gr\"ummer and K.Goeke, {\AP202(1990)297*};
T.Meissner and K.Goeke, \NPA524(1991)719*}
\RF\Grummer88{F.Gr\"ummer, RUB-ITP II/Juny 1988, Preprint Bochum
(unpublished)}
\RF\UMe89{U.-G.Meissner, {\PRL62(1989)1012*};\PLB220(1989)1*}
\RF\Ber87{V.Bernard, Ulf-G.Meissner and I.Zahed, {\PRD36(1987)819*};
{\PRL59(1987)966*}; M.Jaminon, G.Ripka and P.Stassart, \NPA504(1989)733*}
\RF\Walecka74{J.D.Walecka, {\AP83(1974)491*}; B.D.Serot and
J.D.Walecka, Adv. in Nucl. Phys. v.16, eds. J.W.Negele and E.Vogt
(Plenum Press, New York, 1986)}
\RF\Mahaux87{C.Mahaux and R.Sartor, \NPA475(1987)247*}
\RF\Schaldach91{J. Schaldach, E. Thommes, Chr.V.Christov and K.Goeke, DFG
Spring Meeting Salzburg 1992, February 24-28,1992, p.115}
\RF\Jaminon89{M.Jaminon, G.Ripka and P.Stassart, \NPA504(1989)733*}
\RF\Adkins83{G.S.Adkins, C.R.Nappi and E. Witten, \NPB228(1983)552*}
\RF\Goeke91{K.Goeke, A.G\'orski, F.Gr\"ummer, Th.Mei\ss ner, H. Reinhardt
and R. W\"u nsch, \PLB256(1991)321*}
\RF\Gorski92{A.G\'orski, F.Gr\"ummer and K.Goeke, \PLB278(1992)24*}
\RF\TMei91{Th.Meissner and K.Goeke, \ZPA339(1991)513*}
\RF\Kirch91{M. Kirchbach and D.O.Riska, \NCA104(1991)1837*}
\RF\Alberto90{P.Alberto, E.Ruiz Arriola, J.Urbano and
K.Goeke, \PLB247(1990)210*}
\RF\Pobilitza92{P. Pobiltza, E.Ruiz Arriola, F.Gr\"ummer, Th.Mei\ss ner,
K.Goeke and W. Broniowski, \JPG(1992)(in print)*}
\RF\Guad84{A. Guadagnini, \NPA236(1984)35*}
\FIG\Figr1{The energy of the nucleon (thick full line), delta (thin full
line) as well as the non-quantized hedgehog soliton energy(dashed line) as
function of the constituent quark mass.}
\FIG\Figr2{The proton electric form factor as a function of the
squared momentum transfer . The solid lines describe the
form factor and the sea contribution (thin line) for the
constituent quark mass $360\,MeV$. The dashed lines display the same
information for the constituent quark mass $420\,MeV$.}
\FIG\Figr3{The neutron electric form factor as a function of the
squared momentum transfer $Q^2$. The solid lines describe the total
form factor and the sea contribution (thin line) for the
constituent quark mass $360\,MeV$. The dashed lines display the same
information for the constituent quark mass $420\,MeV$.}
\FIG\Figr4{The neutron  charge distribution as a function of the radius.
The solid bold line denotes the total charge, the dashed line denotes the
valence contribution and the thin line stands for the sea contribution
only.}
\FIG\Figr5{Soliton mass as a function of the baryon density for two
different values of the constituent quark mass is compared with the mass of
the soliton (dash-dotted line) of the medium modified $\zs$-model. The
separated contribuions coming from the valence quarks as well as from the
polarization of the Fermi and Dirac sea for the constituent mass $M=420$ MeV
are also presented.}
\FIG\Figr6{Soliton mean square radius as a function of the baryon
density for two different values of the constituent quark mass.}
\FIG\Figr7{The deviation of the theoretical masses from the experimental
values as a function of the strangness quark mass for constituent quark mass
$M$=420 MeV.}
\TAB\Tabl1{The nucleon squared radii computed for the
constituent quark masses. Total, valence and
sea contributions are given separately to see the influence of the
vacuum polarization effects. Listed are also the Coulomb part of the
neutron-proton splitting. In the last column the experimental values
are given for comparison.}
\rthnum{23/92}

\title{Baryons as solitons
in Effective Chiral Quark - Meson Theory{\rm\footnote{$^\ast$}{Talk
presented at the German-Polish Symposium on Particle and Fields, 28.04 -
1.05, 1992, Rydzyna Castle, Poland}}}

\vskip0.5cm
\author{Chr.V.Christov{\rm\footnote{$^{\ddagger)}$}{Permanent
address: Institute for Nuclear Research and Nuclear Energy, Sofia 1784,
Bulgaria}}, A.Blotz, Th. Mei\ss ner, Fr. Gr\"ummer\nl
M. Prasza\l owicz{\rm\footnote{$^{\dagger)}$}{Alexander von
Humboldt Stiftung Fellow, on leave of absence from
Institute of Physics, Jagellonian University, 30-059 Krakow,
Poland}} and K.Goeke}

\address{Institut f\"ur Theoretische Physik II, Ruhr-Universit\"at
Bochum, D-4630 Bochum}

\author{A. Gorski and W. Broniowski}

\address{Institute of Nuclear Physics, 31-342 Krakow, Poland}

\author{D. Dyakonov, V. Petrov and P. Pobylitza}

\address{St.Petersburg Nuclear Physics Institute, Gatchina,
St.Petersburg, Russia}

\vskip0.5cm
\singlespace
\abstract{The Nambu - Jona-Lasinio model in its SU(2) and SU(3) versions
with scalar and pseudoscalar coupling are applied to baryons. The parameters
of the model are fixed in the meson sector. The baryons arise as a
soliton of three valence quarks coupled to the Dirac sea (quark-antiquarks
pairs). Within the SU(2) version the nucleon static properties as well as
some form factors, namely, the electric and  axial ones are described quite
successfully. The nucleon-delta splitting comes out reasonable.
In medium the nucleonic soliton gets less stable -- the mass is reduced
whereas the radius increases. At some critical medium density there is a
clear delocalization of the soliton: The nucleon does not exsist anymore
as a soliton. In SU(3) version the strangness carrying baryons are described
as SU(3)-rotational excitations of the SU(2)-soliton embedded in the
SU(3)-sector. The mass splittings between the octet and decuplet as well
as within the multiplets are peproduced not only in a correct oder but also
in a good agreement with the experimantal values.}

\doublespace
\vskip1cm

\chap{Introduction}

In the last years because of the tremenduos difficulties in the
nonperturbative (strong coupling) regime of QCD there is an increasing
interest in the effective chiral models respecting the chiral symmetry
breaking. Among these models the Nambu -- Jona-Lasinio model applied to
quarks is the only one which allows to combine in a natural way the two
extreme
opposite pictures - the pure valence one advocated by the constituent quark
model and that used in the Skyrme model. The model makes use of the
experimentally supported ``standard'' picture of the nucleon as a bound
state of three valence quarks coupled to the mesons arising as
quark-antiquarks pairs from the Dirac sea. Similar to the other effective
models the NJL model posseses two shortcomings. On the one hand in the model
the quark confinement is not implemented and one can only hope that the
confinement is not of great importanse for the very low-energy baryon
properties. On the other hand the model is not renormalizable. It means
that a finite cut-off is needed to make the theory finite. The cut-off
introduces into the theory a physical scale which should be fixed
by reproducing some physical quantity.

\chap{NJL model with SU(2)-flavour}

We start with the simplest lagrangean of the NJL  model with scalar and
pseudoscalar quark-quark couplings\quref{\Nam61}:
$$
{\cal L}=\overline\Psi i\zg^\zm\partial_\zm\zQ-m_0\overline\zQ\zQ
+\frac G2\lbrack(\overline\zQ\zQ)^2+(\overline\zQ\zg_5\vec\zt\zQ)^2
\rbrack.\EQN\Eq1
$$
Here $\Psi$ describes a quark field with $SU(2)$-flavour ($u$ and $d$)
quarks, $N_c=3$ colours and the average current quark mass
$m_0=(m_u+m_d)/2$. Introducing auxiliary sigma and pion fields by
$\zs=-g\overline\zQ\zQ/\zl^2$ and
${\vec\zp}=-g\overline\zQ i\zg_5\vec\zt\zQ/\zl^2$ and assuming them to
be classical (zero boson and one fermi loop approximation), one can
use path integral techniques in euclidean metric to evaluate the
corresponding effective action as a sum of a quark part
$$
S_{eff}^q=Sp\log\left(\frac \partial{\partial
t}+h-\mu\right)\qquad\hbox{with}\qquad h=\frac {\vec\za.\vec\nabla}i+\zb
g(\zs+i\zg_5\vpi.\vtau) \EQN\Eq3
$$
with $\zm$ being the chemical potential, and a meson one
$$
S_{eff}^m=\int d^4x\left\{\frac
{\zl^2}2(\zs^2+\vpi^2)+\mpi^2\fpi\zs\right\}.
\EQN\Eq2
$$
In the meson part using the PCAC the current mass $m_0$ is eliminated in
favor of the pion mass  $\mpi$ and the pion decay constant $\fpi$.
The coupling constant $G$ is related the new one $\zl$ by $G=g^2/\zl^2$
where the additional coupling constant $g$ is introduced for convinience.
The latter will be fixed later to the physical coupling constant in order
to identify the auxiliary pion and sigma fields with the physical
ones in the meson sector. Only one part of the effective action,
$S_{eff}^q(\zm=0)$, coming from the Dirac sea is divergent. Using the
proper time scheme\quref{\Schwinger51} we make it finite.
Subtracting the vacuum contribution one can express the energy
(see for more details refs.\quref{\TMei90,\Chr90}) as a sum of a quark
part $$
E^q=\sum\limits_{0\le\ze_\za\le\zm}\ze_\za+\frac
{N_c}2\left\{\sum\limits_\za
R_{3/2}(\ze_\za,\zL)-\sum\limits_\za R_{3/2}(\ze^0_\za,\zL)\right\} \EQN\Eq5
$$
with the proper-time regularization function
$$
R_{\za}(\ze,\zL)=\frac 1{\sqrt{4\zp}}\int\limits_{1/\zL^2}^\infty \frac
{d\zt}{\zt^{\za}}\exp({-\ze^2\zt}), \EQN\Eq6
$$
and the meson part which up to a trivial factor is given by
eq.\queq{\Eq2}. The energies $\ze_\za$ are the eigenvalues of the
hamiltonian \queq{\Eq3} and $\ze_\za^0$ correspond to the vacuum solution.

We fix the parmeters of the model, namely the coupling constant $\zl^2$ and
the cut-off $\zL$, in the meson sector of the vacuum by reproducing the
experimental values of the pion decay constant $\fpi$ and the pion mass
$\mpi$. This procedure originates from Eguchi\quref{\Egu76} and
is described in detail in ref.\quref{\TMei90}. We use the stationary meson
field configuration $<\vpi>_v=\vpi_v=0$ and $<\zs>_v\neq 0$ where the
the latter is given by the non-trivial solution
of the well-known gap equation. It generates the constituent quark mass
$M=g<\zs>_v=g\fpi$. Actually we use the gap equation to express the
coupling constant $\zl^2$ as a function of the constituent mass $M$. For the
meson dynamics one can use the meson propagators
\quref{\Jaminon89,\Schaldach91} and to fix the $g$ to the physical pion
coupling constant in order to identify the auxiliary $\zs$ and $\vpi$ with
the physical ones. The next step is to reproduce the physical value
of the pion decay constant. The latter leads to a relation between the $M$
and $\zL$. However, in the present case one can also use the
gradient expansion which means that the coupling constant is fixed to
yield the correct form of the meson kinetic energy term and the
pion mass is related to the second derivative of the energy at the
stationary point. It should be also noticed that in this limit ($q^2=0$)
the pion decay constant is reproduced trivially for the physical pion
fields.
All those conditions together leave the vacuum value of the constituent
mass $M$ as the only free parameter. In principle, the empirical
values\quref{\Shifman79} of the quark condensate and the quark bare mass
can be used to fix $M$ but actually they still leave a broad range for $M$.

For the solitonic sector we solve the Dirac equation (eq.\queq{\Eq2}) with
the equations of motion of the meson fields:
$$
\zs=\frac g{\zl^2}N_c\left\{ \sum\limits_\za \overline\phi_{\za}\phi_{\za}
R_{1/2}(\ze_{\za},\zL)-\sum\limits_{0\le\ze_\za\le\zm}
\overline\phi_{\za}\phi_{\za}\right\}-\frac{\fpi\mpi^2}{\zl^2}. \EQN\Eq8
$$
$$
\zp=\frac g{\zl^2}N_c\left\{
\sum\limits_\za
\overline\phi_{\za}\imath\zg_5(\vec\zt.\hat r)\phi_{\za}
R_{1/2}(\ze_{\za},\zL)-\sum\limits_{0\le\ze_\za\le\zm}
\overline\phi_{\za} \imath\zg_5(\vec\zt.\hat r)\phi_{\za}\right\}. \EQN\Eq9
$$
The meson fields are assumed to be in a hedgehog form and are restricted
on the chiral circle $\zs^2+\vpi^2=\fpi^2$. We use a numerical
self-consistent iterative procedure\quref{\TMei90} based on a method
proposed by Ripka and Kahana\quref{\Ripka84}.

As a next step the classical soliton should be
quantized. To that end one can make use of the rotational symmetry of the
solution introducing isorotating hedgehog meson fields
$$
\widetilde U(x) \rightarrow R(t)U(x)R^\dagger(t), \EQN\Eq9a
$$
where $R(t)$ is a unitary SU(2) matrix. Following the usual canonical
quantization procedure\quref{\Adkins83} one can relate the angular
velocities to the spin (isospin) operators by
$$
\theta \zW^i(\zW^a) \rightarrow -iI^i(T^a).  \EQN\Eq9b
$$
Thus one can assign proper spin and isospin quantum numbers to the soliton
and evaluate the nucleon properties like energies, radii, form factors,
spin and isospin contents, etc.

For the energy an expansion up to the first non-vanishing order with
respect to angular velocity gives the well-known rotational energy
correction to the classical energy from which one should subtract the
corresponding translational and rotational spurious zero-point
energies\quref{\Pobilitza92}
$$
E^I=E_{sol}+\frac{I(I+1)}{2\theta}-\frac {<T^2>}{2\theta}-\frac
{<P^2>}{2\theta} . \EQN\Eq10
$$
Here $\theta$ is the moment of inertia given as a sum of valence and sea
part\quref{\Goeke91}. The latter is reguralized in the proper time scheme.
The results of the energy of the nucleon and the delta are shown in
\qufig{\Figr1} and some values of the moment of inertia for different
constituent masses are presented in \qutab{\Tabl1}. As can be seen the
experimental nucleon-delta mass splitting is reproduced well for $M$ around
420 - 450 MeV. The sea contribution is less than 20 \%.

The nucleon electric form factors are given by
$$
G_E(q^2) =\bra {N(p)}\it j^{em}_o(0)\ket{N(p')} , \EQN\Eq11
$$
where $\ket {N(p)}$ is the quantized nucleon state and  $\it j^{em}_o$ is
zero-component of the electromagnetic current operator defined as
$$
\it j^{em}_\zm=\frac \zd{A_\zm}S^f_{eff}[A_\zm,\zW].  \EQN\Eq12
$$
To that end an external electromagnetic field $A_\zm$ is coupled. Similar to
the moment of inertia the form factors incude also a valence and
sea contribution. It should be noted that the isoscalar form factor does
not need any regularization which is not the case of the isovector one. The
proton and neutron electric form factors calculated\quref{\Gorski92} in the
non-relativistic
approximation $q^2\ll M_N^2$ are shown in \qufig{\Figr2} and \qufig{\Figr3},
respectively. The corresponding neutron charge distribution presented in
\qufig{\Figr5} has the typical negative tail coming from the pion field
(sea contribution). The proton form factor is in good agreement with the
experiment whereas the neutron one is overestimated by a factor two. Both
form factors show practically no dependence on the constituent mass $M$.
For the proton the sea contribution is less than 5\% which is not the case
of the neutron one where the valence quarks and the sea contribute with
similar magnitude and opposite signs. Concerning the neutron electric form
factor one could expect that both the boson loop corrections
as well as the incorporation of vector
mesons\quref{\Alberto90} may modify it. As can be expected the
isoscalar squared radius is reproduced well whereas the isovector one is by
a factor of two too large (see \qutab{\Tabl1}). The \qutab{\Tabl1} lists the
Coulomb part of the proton-neutron mass splitting as well.

In non-relativistic limit the axial form factor defined as usual by
$$
\bra {N_2(p)}A^a_i(0)\ket{N_1(p')}=g_A(q^2)\left(\zd_{ij}-\frac{q_iq_j}{\mid
q\mid^2}\right) \bra {\chi_2}\zs_i\zt^a \ket{\chi_1} \EQN\Eq13
$$
is also calculated\quref{\TMei91} using the quantized nucleon state
$\ket{N_1(p')}$. As can be seen from \qufig{\Figr5} the results are
in good agreement with the experimantal dipole fit at finite momentum
transfer $Q^2$. The value at the origin (about 0.8),
which is exactly the axial vector coupling constant, is lower than the
experimental value 1.2. However, it is a common problem of the effective
chiral models. Kirchbach and Riska\quref{\Kirch91} relate it to the problem
of a proper renormalization of the axial charge.

\singlespace
\item{Table 1} The nucleon squared radii computed for the
constituent quark masses. Total, valence and
sea contributions are given separately to see the influence of the
vacuum polarization effects. Listed are also
the $N-\zD$ mass splitting as well as the Coulomb part of the
neutron-proton mass splitting. In the last column the experimental values
are given for comparison.

\vskip0.5cm

\tenpoint
\vbox{\offinterlineskip
\hrule height1pt
\halign{&\vrule width1pt#&
 \strut\quad\hfil#\quad
 &\vrule#&\quad\hfil#\quad
 &\vrule#&\quad\hfil#\quad
 &\vrule#&\quad\hfil#\quad
 &\vrule#&\quad\hfil#\quad
 &\vrule#&\quad\hfil#\quad
 &\vrule#&\quad\hfil#\quad
 &\vrule#&\quad\hfil#\quad
 &\vrule width1pt#&\quad\hfil#\quad\cr
height6pt&\omit&&\multispan{11}&&\omit&\cr
& &&\multispan{11} \hfill {\bf Constituent\ Quark\ Mass} \hfill &&
&\cr
height4pt&\omit&&\multispan{11}&&\omit&\cr
&\hfill {\bf Quantity}\hfill &&\multispan3 \hfill 320\ MeV\hfill
&&\multispan3
\hfill 420   MeV\hfill
&&\multispan3
\hfill 465   MeV\hfill
&&\hfill {\bf Experiment} &\cr
height3pt&\omit&&\multispan3 && \multispan3 &&
\multispan3 &&\omit&\cr
\noalign{\moveright 37.8mm \vbox{\hrule width98mm}}
height3pt&\omit&&\omit&&\omit&&\omit&&\omit&&\omit&&\omit&&\omit&\cr
&\hfill    &&\hfill total &&\hfill sea &&\hfill total &&\hfill sea &&
\hfill total &&\hfill sea &&   &\cr
height3pt&\omit&&\omit&&\omit&&\omit&&\omit&&\omit&&\omit&&\omit&\cr
\noalign{\hrule height1pt}
height3pt&\omit&&\omit&&\omit&&\omit&&\omit&&\omit&&\omit&&\omit&\cr

& \hfill $ <r^2>_{T=0} \ \ [fm^2] $ &&
\hfill 0.66  &&
\hfill 0.07  &&
\hfill 0.61  &&
\hfill 0.16  &&
\hfill 0.53  &&
\hfill 0.15  &&
\hfill 0.62\hfill
&\cr
height3pt&\omit&&\omit&&\omit&&\omit&&\omit&&\omit&&\omit&&\omit&\cr
\noalign{\hrule}
height3pt&\omit&&\omit&&\omit&&\omit&&\omit&&\omit&&\omit&&\omit&\cr

& \hfill $ <r^2>_{T=1} \ \ [fm^2] $ &&
\hfill 1.36  &&
\hfill 0.71  &&
\hfill 1.15  &&
\hfill 0.74  &&
\hfill 1.05  &&
\hfill 0.73  &&
\hfill 0.86\hfill
&\cr
height3pt&\omit&&\omit&&\omit&&\omit&&\omit&&\omit&&\omit&&\omit&\cr
\noalign{\hrule}
height3pt&\omit&&\omit&&\omit&&\omit&&\omit&&\omit&&\omit&&\omit&\cr

& \hfill $ <r^2>_n\ \ [fm^2] $ &&
\hfill --0.35  &&
\hfill --0.32  &&
\hfill --0.27  &&
\hfill --0.29  &&
\hfill --0.26  &&
\hfill --0.29  &&
\hfill --0.12 \hfill&\cr
height3pt&\omit&&\omit&&\omit&&\omit&&\omit&&\omit&&\omit&&\omit&\cr
\noalign{\hrule}
height3pt&\omit&&\omit&&\omit&&\omit&&\omit&&\omit&&\omit&&\omit&\cr

& \hfill $ <r^2>_p \ \ [fm^2] $ &&
\hfill 1.01  &&
\hfill 0.39  &&
\hfill 0.88  &&
\hfill 0.45  &&
\hfill 0.79  &&
\hfill 0.44  &&
\hfill 0.74\hfill&\cr
height3pt&\omit&&\omit&&\omit&&\omit&&\omit&&\omit&&\omit&&\omit&\cr
\noalign{\hrule}
height3pt&\omit&&\omit&&\omit&&\omit&&\omit&&\omit&&\omit&&\omit&\cr

& \hfill $ M_{\Delta}-M_N \ \ [MeV] $ &&
\hfill 203  &&
\hfill ---\hfill  &&
\hfill 261\hfill &&
\hfill ---  &&
\hfill 301\hfill  &&
\hfill ---\hfill &&
\hfill \ 294 \hfill&\cr
height3pt&\omit&&\omit&&\omit&&\omit&&\omit&&\omit&&\omit&&\omit&\cr
\noalign{\hrule}
height3pt&\omit&&\omit&&\omit&&\omit&&\omit&&\omit&&\omit&&\omit&\cr

& \hfill $ \Delta_{EM}\ \ \ \ [MeV] $\hfill &&
\hfill --0.87  &&
\hfill --0.18  &&
\hfill --1.00  &&
\hfill --0.73  &&
\hfill --1.09  &&
\hfill --0.33  &&
\hfill --0.76 $\pm$0.30 \ \hfill &\cr
height3pt&\omit&&\omit&&\omit&&\omit&&\omit&&\omit&&\omit&&\omit
&\cr
\noalign{\hrule height1pt} }}

\doublespace
\twelvepoint
\vskip0.5cm

The sigma commutator calculated in the model comes out quite reasonable -
about 35 MeV.

We also studied possible modifications of the nucleon properties in medium
(finite density). The medium is simulated simply by varying the chemical
potential $\zm$ in order to fix the medium density. The mass of the
nucleonic soliton is depicted in \qufig{\Figr6} as function of the density.
With increasing density the mass gets reduced as at some critical value the
soliton disappears - a sign of a delocalization of the nucleon. The mean
radius shown in \qufig{\Figr6} supports this picture. At finite density the
soliton gets extended (swelling) and less bound and close to some critical
density it is getting delocalized.

\chap{NJL model with SU(3)-flavours}

The SU(3)-invariant NJl model withscalar and pseudoscalar coupling reads
$$
{\cal L}_{NJL}={\bar q}(x) ( i \dsl - m)q(x) - {G\over 2}
 \left[ ({\bar q}(x)\lambda^aq(x))^2 +
	 ({\bar q}(x)i\gamma_5 \lambda^a q(x))^2  \right]
 \EQN\Eq14
$$
where $m={\rm diag}(m_u,m_d,m_s)=m_1 {\bf 1 }
+m_3\lambda_8$   is
the quark mass matrix and $\lambda_a$ are the usual Gell-Mann matrices
with $\lambda^0=\sqrt{(\frac 23)} {\bf 1}$. Actually we use the current
quark masses  $m_0=\half(m_u+m_d)$. Integrating out the quarks the effective
action is given by
$$  S_{eff} =  -{\rm Sp \ \log }  \left( i \dsl - m
       - g  (\sigma^a \lambda^a + i \gamma_5 \pi^a \lambda^a) \right)
    +{\mu^2 \over 2} (\sigma^a \sigma^a + \pi^a \pi^a  )
    \EQN\Eq15 $$
Hereafter we work in the chiral limit and treat the strange quarkmass
$m_s$ perturbatively in first  order. Similar to the SU(2) case
we fix the parameters in the meson sector of the vacuum by reproducing the
experimental values of the meson masses $m_\pi=139MeV$ and $m_K=496MeV$,
as well as the pion decay constant $f_\pi=93MeV$.
This results in the relation ${m_K^2/m_\pi^2}={(m_s+m_0)/2m_0}$ with
$m_0=\half(m_u+m_d)$. In order to fulfill it one needs a current mass
$m_0$ of about 6 MeV. To this end we use a proper time
regularization  with a modified weight function. The latter
leads to a higher value of the $\zS$-commutator $\zS=51$ MeV.

For the SU(3) soliton we use the $SU(2)$-one $U_0$ trivially embedded in
SU(3) $$
U_2=\mat{&U_0&0\cr&0&1}\EQN\Eq16
$$
In order to quantize the classical soliton we follow the same scheme
(cranking approxaimation) as in the SU(2) case (see eq.\queq{\Eq9a}) but
with a SU(3) isorotating meson fields. As a next step we expand the
effective action in $\Omega_A$ up to second order
$$
L^{rot}= \frac 12 I_{AB} \Omega_A \Omega_B -{N_c \over 2 \sqrt{3}  }
 \Omega_8, \EQN\Eq17
$$
where $I_{AB}$ is the $SU(3)$ tensor of the moment of inertia,
as well as up to the first order in
$m_s$
$$ L^{(1)} = -{2\over 3} {m_s \over m_u +m_d} \Sigma
\left( 1 - D_{88}^{(8)}(R)\right)
   \EQN\Eq18
$$
and $m_s\Omega_A$
$$
L^{(2)}=-{2m_s\over \sqrt{3} } K_{AB} D_{8A}^{(8)}(R)\Omega_B.
     \EQN\Eq19
$$
All other corrections are neglected.
Here $D_{8A}^{(8)}(R)$ is the SU(3)-Wigner function and by $K_{AB}$ we
denote the tensor of the anomalous moments of inertia.
These moments need no regularization, because they originate from the
imaginary part of the effective Euclidean action and so they are finite.
It should be stressed that the appearance of the linear term in $\Omega_8$
in \queq{\Eq17 } is due to the discrete valence level in the spectrum
whereas in the Skyrme model a term like this can be obtained only by adding
the Wess-Zumino term.

Now we quantize canonically by defining $SU(3)$ right generators
$R_{A=1,...,8}$ by  $R_A=-{{\partial L}/{\partial\Omega_A}}$,
where $L=L^{rot}+L^{(1)}+ L^{(2)}$.
Using the collective hamiltonian, obtained in approximation ${\cal
O}(m_s^2)$, in a strict perturbation theory to first order in  $m_s$ we
calculated the masses of the member of the octet and decouplet.
We fixed $m_{\Sigma^*}$ to the experimental value, because
as in the $SU(2)$-case  the total masses  are always  too large by a
constant shift (zero-point energies are not subtracted). It should be
stressed that we obtained the Gell-Mann Okubo relations
$$
2(m_N+m_{\Xi})=3m_\Lambda+m_\Sigma\qquad\hbox{and}\qquad
m_\Omega-m_{\Xi^*}=m_{\Xi^*}-m_{\Sigma^*}=m_{\Sigma^*}-m_{\Delta}
$$
as well as the Guadignini formula\quref{\Guad84}
$$
m_{\Xi^*}-m_{\Sigma^*}+m_N=(1/8)(11m_\Lambda-3m_\Sigma)
$$
without any
additional assumptions. The latter is derived in the Skyrme model only by
adding a term associated to the hypercharge simply by hand. Our results are
illustrated in \qufig{\Figr7} where the discrepancy between the theory and
experiment for the masses of the members of the octet and decouplet is
presented as a function of the strangness quark mass.

\chap{Summary}

The SU(2)-flavour Nambu -- Jona-Lasinio model with scalar and
pseudoscalar couplings and valence quarks combined with the semi-classical
quantization (cranking approximation) provides quite a reasonable
description of nucleon properties like masses, charge radii as well as
electric and axial form factors. In the SU(3)-version the mass splittings
between the octet and decuplet as well as within the multiplets are
peproduced not only in a correct oder but also in a good agreement with the
experimantal values. Thus, the assumed physical picture of the baryon
as a soliton of three valence quarks coupled to the Dirac sea is supported.
The results do not leave much room for additional meson degrees of freedom,
in particular for vector mesons. Since the NJL model respects the chiral
symmetry breaking but not confinement one can conclude that i) the
chiral symmetry breaking mechanism is dominant for the nucleon structure
and ii) the confinement seems to be not of great importance at least for the
very low-energy properties considered. In particular, the (partial)
restoration of the chiral symmetry occuring in medium modifies
of the nucleon structure: the nucleon gets less bound and swelled and at
some critical density it undergoes delocalization.

{\it The work has been supported by the Bundesministerium f\"ur
Forschung und Technologie, Bonn, by the KFA J\"ulich (COSY -- Project),
the Deutsche Forschunggemeinschaft and by Polish Research Grant
2-0091-91-01.}

\refout
\figout
\end